\documentclass[preprint,aps]{revtex4}

\usepackage{graphicx}
\usepackage{latexsym,amssymb,amsmath,wasysym}

\begin{document}


\title{Velocity Correlations in Dense Gravity Driven Granular Chute Flow}

\newcommand{\av}[1]{\langle #1 \rangle}
\newcommand \ts {\delta t} 
\newcommand \dt {\Delta t} 
\newcommand \vd {v_\alpha} 
\newcommand \mv {\bar{v}_\alpha} 
\newcommand \cyy {C_{yy}} 
\newcommand \gd {\dot{\gamma}} 
\newcommand \lv {l_{\nu}} 
\newcommand \dg {^\circ} 
\newcommand{\hstop}{h_{\rm stop}}  
\newcommand{\lab}{\lambda_{\alpha\beta}} 
\newcommand{\lyy}{\lambda_{yy}} 
\newcommand{\Cab}{C_{\alpha\beta}} 
\newcommand{\Lab}{L_{\alpha\beta}} 
\newcommand{\zlayer}{z_\textrm{layer}}  
\newcommand{\Nl}{N_\textrm{layer}}  

\author{Oleh Baran}
\affiliation{Corporate Strategic Research, ExxonMobil Research and
  Engineering, Annandale, New Jersey 08801}

\author{Deniz Erta\c{s}}
\affiliation{Corporate Strategic Research, ExxonMobil Research and
  Engineering, Annandale, New Jersey 08801}

\author{Gary S. Grest}
\affiliation{Sandia National Laboratories, Albuquerque, New Mexico 87185}

\author{Thomas C. Halsey}
\affiliation{Corporate Strategic Research, ExxonMobil Research and
  Engineering, Annandale, New Jersey 08801}

\author{Jeremy B. Lechman}
\affiliation{Sandia National Laboratories, Albuquerque, New Mexico 87185}

\date \today

\begin{abstract}

We report numerical results for velocity correlations in dense,
gravity-driven granular flow down an inclined plane.  For the grains
on the surface layer, our results are consistent with experimental
measurements reported by Pouliquen. We show that the correlation
structure within planes parallel to the surface persists in the
bulk. The two-point velocity correlation function exhibits exponential
decay for small to intermediate values of the separation between
spheres. The correlation lengths identified by exponential fits to the
data show nontrivial dependence on the averaging time $\dt$ used to
determine grain velocities. We discuss the correlation length
dependence on averaging time, incline angle, pile height, depth of the
layer, system size and grain stiffness, and relate the results to
other length scales associated with the rheology of the system. We
find that correlation lengths are typically quite small, of the order
of a particle diameter, and increase approximately logarithmically
with a minimum pile height for which flow is possible, $\hstop$,
contrary to the theoretical expectation of a proportional relationship
between the two length scales.
\end{abstract}

\maketitle

\section{Introduction}

Dense, gravity-driven granular flows down an inclined
plane~\cite{jop06,silbert01,lemieux85,azanza99} achieved the status of
a key model system because of their relevance to many geological and
industrial applications. Provided that the surface of the incline is
sufficiently rough and the flow height is small compared to the width
and length of the chute, such that transients and side-wall effects
can be neglected, the flow behavior is controlled by the angle of
inclination, $\theta$, and granular layer thickness $h$.  For a given
$\theta$ the flow is possible only for $h>\hstop(\theta)$. Here
$\hstop$ is often referred to as the deposit function, because it is
approximately equal to the thickness of the deposit remaining on the
plane once the flow is stopped either due to the decrease of $\theta$
or decrease of $h$. Recent experimental and numerical studies
confirmed that for spherical grains, the depth-averaged steady state
velocity of the flow $u$ can be related to the deposit function
through the following relationship, originally proposed by
Pouliquen~\cite{pouliquen99,silbert01}:
\begin{equation}
\frac{u}{\sqrt{gh}} = \beta \frac{h}{\hstop(\theta)},
\label{eq:hstop}
\end{equation}
where $g$ is the gravitational acceleration and $\beta\approx0.13$. 
This relationship suggests that a single scaling length controls both
the deposit function and the rheology.

A potential explanation of this scaling length as arising from
correlations in granular motion was advanced by some of
us~\cite{ertas02}, linking a correlation length in the flow to a
rheologically defined ``viscosity length scale'', $l_{\nu}$, defined
by the Bagnold scaling form~\cite{ertas02}
\begin{equation}
\sigma_{xz}=\rho l_{\nu}^2\gd^2,
\label{eq:l_n}
\end{equation}
for a granular system with bulk density $\rho$, flowing under a shear
stress $\sigma_{xz}$ at a shear rate $\gd$.  Alternatively, it has
recently been argued~\cite{midi04,cruz05} that for a granular system
made of particles of diameter $d$ and grain density $\rho_g$ flowing
while subject to a pressure $P$, the rheology is controlled by the
local scaling variable
\begin{equation}
I = \frac{\gd d}{\sqrt{P/\rho_g}} ,
\label{eq:gdrm}
\end{equation}
which is a generalization of Eq.(\ref{eq:hstop}). For steady state
chute flow, these two quantities are related through
\begin{equation}
I=\frac{d}{l_{\nu}}\sqrt{\frac{\tan\theta}{\phi}},
\label{eq:Ivsl_n}
\end{equation}
whre $\phi$ is the volume fraction, thus they can be used
interchangeably in this case. The fact that the same rheology can be
advanced with or without an assumed fundamental length scale (a
``correlation length'') raises questions about the correct
interpretation of this length scale, and the nature of its
relationship to flow-induced structures, if
any. Pouliquen~\cite{pouliquen04} has recently reported experimental
measurements of two point velocity correlation functions at the flow
surface, and obtained results supporting a possible connection between
the observed correlation length and the deposit function.

In this work, we repeat and extend the correlation analysis presented
by Pouliquen~\cite{pouliquen04} to layers that are far from the bottom
and surface of the flow, where true bulk rheology is observed, by
taking advantage of information available from discrete element (DEM)
simulations that is difficult to obtain experimentally.  We are able
to reproduce and further analyze the experimental results, and
ultimately make the following observations:

\begin{itemize}
  \item{All two-point velocity correlation functions exhibit exponential
decay with relative distance, generally with different values of the 
correlation length $\lab$ for different components $\beta$ of the velocity 
and $\alpha$ of the displacement vector.}
  
  \item{The measured correlation lengths $\lab$ vary with the velocity
measurement time $\dt$ used to calculate the velocities from
displacements. In order to be able to compare correlation lengths at
different depths and between different runs, it is necessary to adjust
the measurement time such that the particles that are being measured
experience a fixed, predetermined strain $\epsilon = \gd \dt$. This
yields correlation lengths that are independent of the position of the
measured layer in the bulk, or the overall flow height.}

  \item{Correlation lengths in the bulk are uniquely determined by the incline
angle $\theta$ (or equivalently by $\hstop$ or $I$). They increase with
decreasing $\theta$.}

\item{Correlation lengths are typically quite small, of the order of
a particle diameter, and increase approximately logarithmically 
with $\hstop$, contrary to the theoretical expectation of a proportional
relationship between the two length scales.}

\item{For very high piles and angles close to the angle of repose, the
flows excite a low frequency ``breathing mode'' with coherent motion
normal to the surface layer, previously observed by
Silbert~\cite{silbert05}. This limits our ability to probe the limit
of large $\hstop$, small $I$ in this geometry.}

\end{itemize}

The paper is organized as follows: In Section~\ref{sec:method} we
outline the simulation method, which has been described in detail
elsewhere~\cite{silbert01}, and explain the analysis used to obtain
the reported results. In Section~\ref{sec:surface} we present the
results of velocity correlations analysis in the surface layer and
compare these results with those of Pouliquen~\cite{pouliquen04}. In
Section~\ref{sec:depth} we discuss the differences between
correlations on the surface and in the bulk. We then report the
results for velocity correlations in the bulk in
Section~\ref{sec:bulk}.  We also discuss the velocity fluctuations in
Appendix~\ref{sec:fluc} and the effect of system size and stiffness of
particles on the correlation length in
Appendix~\ref{sec:large}. Finally, in Section~\ref{sec:conclusion} we
present our conclusions.

\section{\label{sec:method} Simulation and Analysis Method}
\begin{figure}[!ht]
\includegraphics[width=8cm]{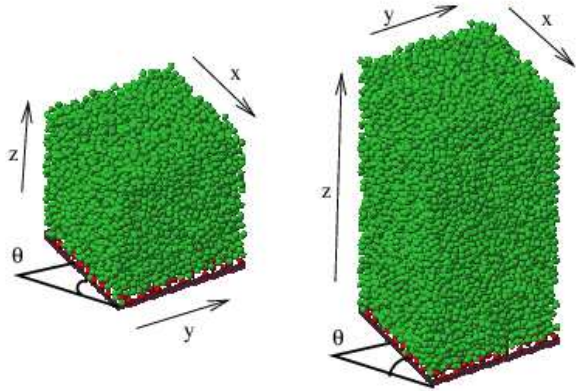}
\caption{\label{fig:snapshot} (Color online) Typical snapshots from
simulations with labeled coordinate axes. $\theta$ is an angle of
inclination of the rough base. Base area $20d\times 20d$. 9089 spheres
in the system of height $h/d\simeq 20$ (left) and 17889 spheres in the
system of height $h/d\simeq 40$ (right).}
\end{figure}

\subsection{Simulation Method}

We perform three-dimensional discrete element (DEM) simulations of
spherical, monodisperse particles of diameter $d$ in the chute flow
geometry.  The simulation method is described in detail in
Ref.~\cite{silbert01}.  The simulation cell consists of a rough bottom
in the $xy-$ plane, whose normal is tilted by an angle $\theta$ with
respect to the direction of gravitational acceleration $g$ in order to
induce a flow in the $x-$direction, as shown in
Fig.~\ref{fig:snapshot}. The bottom is roughened by a thin substrate
of stationary particles uniformly distributed over the base of the
system with the same properties as the mobile particles. Periodic
boundary conditions are imposed in $x-$ and $y-$ directions with a
typical base size of $20d\times 20d$, in order to eliminate the
documented effect of side walls~\cite{taberlet03}.

The mobile particles flowing above the base are spheres of mass $m$
and diameter $d$ that form Hertzian contacts as well as interacting
frictionally. Unless otherwise noted, the simulation parameters used
for the runs were $k_n= 2\times 10^5 \, mg/d$ (normal contact
stiffness), $\mu=0.5$ (Coulomb friction coefficient), $k_t=(2/7)k_n$
(tangential contact stiffness), $\gamma_n = 50 \, \sqrt{g/d}$ (normal
viscoelastic constant) and $\gamma_t=0$ (tangential viscoelastic
constant)~\footnote{The contact stiffness is related to the Young's
modulus $E$ and Poisson ratio $\nu$ of the spheres through
$k_n=\frac{Ed}{3(1-\nu^2)}$. The typical value of stiffness
$k_n=2\times 10^5 \, mg/d$ used in our simulations corresponds
approximately to the behavior of $1$ cm diameter rubber balls. As is
shown in previous simulations~\cite{silbert01} and in
Appendix~\ref{sec:large}, increasing $k_n$ does not change any result
qualitatively.}. In Appendix~\ref{sec:large} we also present results
with reduced ($k_n=2\times 10^3\, mg/d$) and increased ($k_n=2\times
10^7\, mg/d$) stiffness of particles.

We evolve the system by integrating the appropriate equations of
motion~\cite{silbert01} with a timestep $\ts /\tau_0 = 10^{-4}$, where
the characteristic time $\tau_0=\sqrt{d/g}$.  In order to ensure
measurements in steady state, we equilibrate each system for a long
enough time ($1-2\times 10^7$ $\ts$) and check the stationarity of
velocity profiles. Our main comparative analysis is applied to
simulations with a base area of $20d\times 20d$ with two different
flow heights ($h/d \simeq 20$ with 9089 particles and $40$ with 17889
particles) at different angles $\theta$ in the range between $20.5\dg$
and $26\dg$.

\begin{figure}[!ht]
\includegraphics{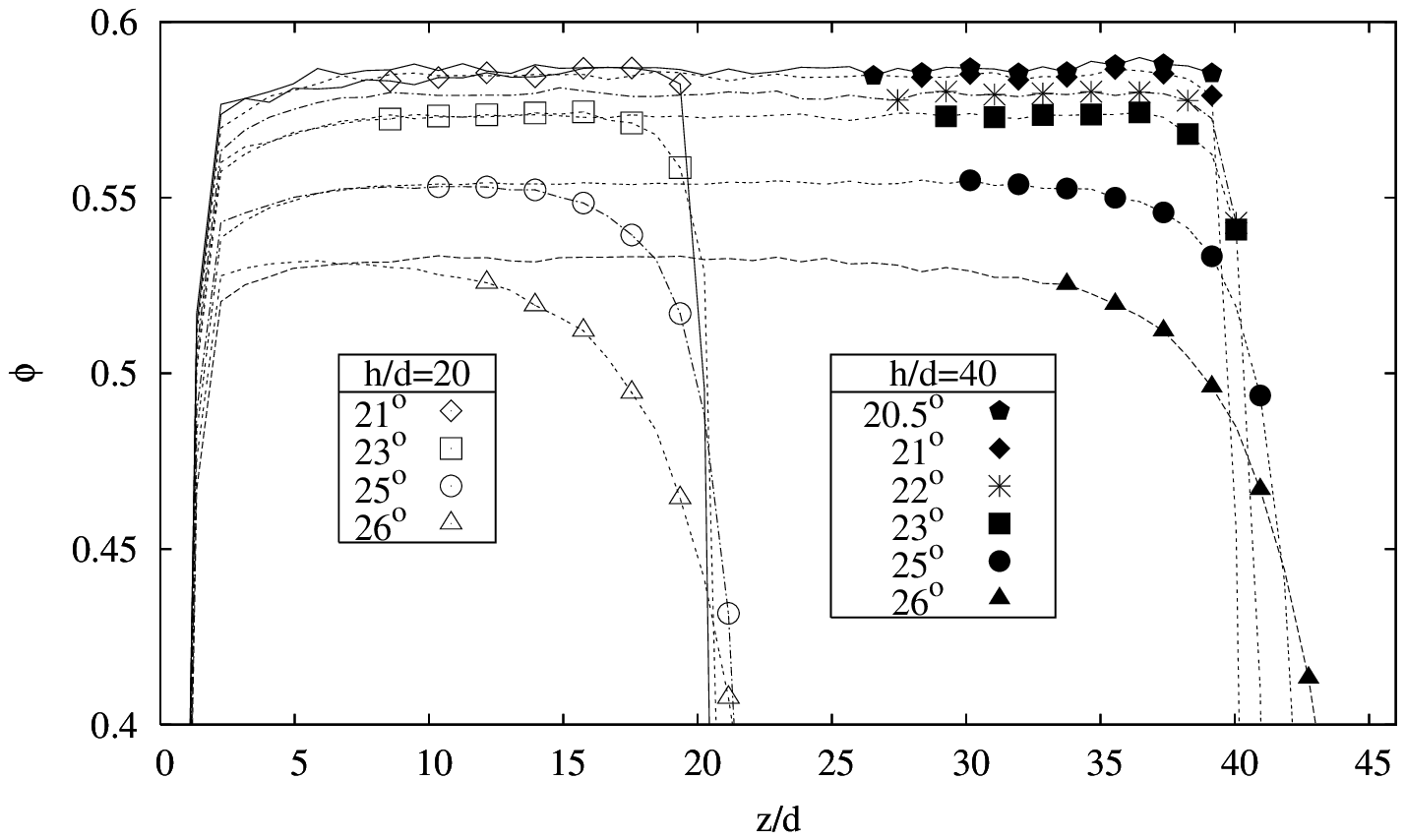}
\includegraphics{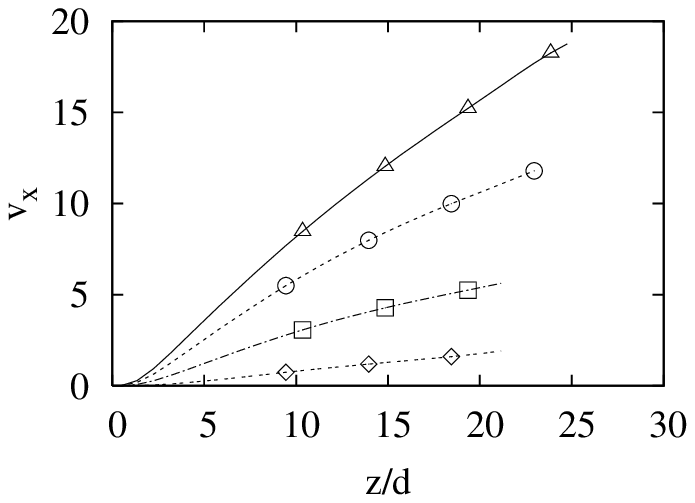}
\includegraphics{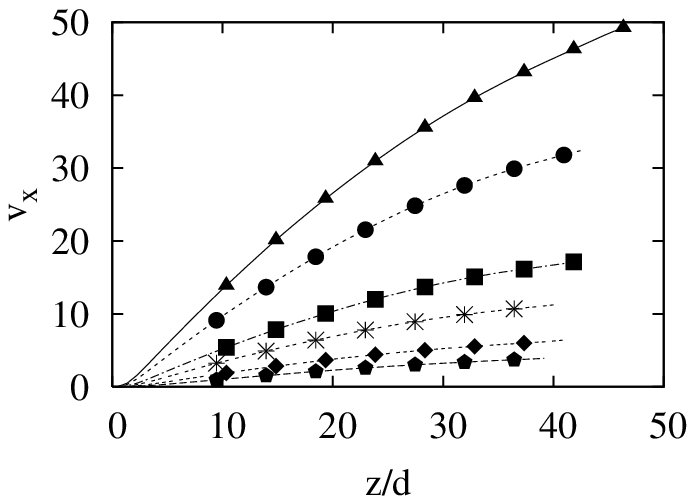}
\caption{\label{fig:prof} (a) Volume fraction profiles for all systems
  listed in Table~\ref{tab:layers_part}. (b) Velocity profiles for
  $h/d \simeq 20$. (c) Velocity profiles for $h/d \simeq 40$. See also
  Table~\ref{tab:layers_part} for legend information. All profiles are
  obtained with bin size of $0.9d$. All data points are connected with
  lines and the symbols are used to mark each data-set.}
\end{figure}
The main properties of the runs with typical parameters, as well as
the results of the correlation analysis for the selected layers in the
bulk of these flows, are summarized in Table~\ref{tab:layers_part}.
The steady-state volume fraction and velocity profiles of these runs
are depicted in Fig.~\ref{fig:prof}. In addition, we simulated and
obtained selective results for thin piles, $h/d \simeq 10$, with
typical ($20d \times 20d$) and large ($40d \times 40d$) base area (see
Appendix~\ref{sec:large}) as well as taller piles, $h/d \simeq 80$.

\begin{table*}[!ht] 
\begin{ruledtabular}
\begin{tabular}{|c||c|c|c|c|c|c|c|c|c|c|}

$\theta$ & $20.5\dg $  & $21\dg $  & $21\dg $  & $22\dg $  & $23\dg $  & $23\dg $  & $25\dg $  & $25\dg $  & $26\dg $  & $26\dg $   \\ \hline

$\hstop/d^{\dag}$ & $19.6$  & $16.4$  & $16.4$  & $12.1$  & $9.0$  & $9.0$  & $5.0$  & $5.0$  & $3.6$  & $3.6$   \\ \hline

$h/d$ & $40$  & $20$  & $40$  & $40$  & $20$  & $40$  & $20$  & $40$  & $20$  & $40$   \\ \hline

Legend & $\includegraphics{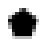}$  & $\lozenge$  & $\blacklozenge$  & $\ast$  & $\square$  & $\blacksquare$  & $\bigcirc$  & $\CIRCLE$  & $\triangle$  & $\blacktriangle$   \\ \hline

$\phi_\textrm{bulk}$ & $.586$  & $.584$  & $.584$  & $.578$  & $.573$  & $.572$  & $.553$  & $.554$  & $.526$  & $.532$   \\ \hline \hline

$z_\textrm{layer}/d^\ddag$ & $25.0$  & $11.5$  & $25.0$  & $25.0$  & $11.5$  & $25.0$  & $11.5$  & $25.0$  & $11.5$  & $25.0$   \\ \hline

$\av{v_x}$ & $2.79$  & $.95$  & $4.57$  & $8.32$  & $3.40$  & $12.46$  & $6.68$  & $23.09$  & $9.42$  & $32.17$   \\ \hline

$\gd\tau_0$ & $.096$  & $.101$  & $.144$  & $.258$  & $.286$  & $.390$  & $.556$  & $.730$  & $.820$  & $1.053$   \\ \hline

$I $ & $.025$  & $.034$  & $.038$  & $.067$  & $.096$  & $.101$  & $.182$  & $.183$  & $.259$  & $.253$   \\ \hline

$\av{(u_x^{inst})^2} $ & $.018$  & $.014$  & $.026$  & $.060$  & $.060$  & $.109$  & $.186$  & $.314$  & $.395$  & $.639$   \\ \hline

$\av{(u_y^{inst})^2} $ & $.011$  & $.010$  & $.020$  & $.046$  & $.046$  & $.084$  & $.139$  & $.234$  & $.296$  & $.471$   \\ \hline

$\av{(u_z^{inst})^2} $ & $.019$  & $.012$  & $.026$  & $.055$  & $.055$  & $.098$  & $.162$  & $.272$  & $.338$  & $.539$    \\ \hline

$\lambda_{yy}^{\epsilon=0.1}/d$ & $1.36$  & $1.40$  & $1.34$  & $1.24$  & $1.18$  & $1.17$  & $1.02$  & $1.00$  & $0.84$  & $0.86$   

\end{tabular}
\end{ruledtabular}
\caption{\label{tab:layers_part} Properties of the simulation runs and 
associated bulk layers that were analyzed for velocity correlations.\\
$\dag$ : From Ref.~\cite{silbert03}: The tangential viscoelastic constant
$\gamma_t=\gamma_n/2$. However, $\hstop$ is not affected 
by this difference.\\
$\ddag$ : Layer thickness is one diameter, centered around the 
reported $z_\textrm{layer}$.}

\end{table*}

\subsection{\label{sec:analysis}Velocity Correlation Analysis Method}

All correlation analyses are performed at steady state, within thin
layers normal to the $z-$axis. The surface correlation analysis was
applied to a preset number of particles ($\Nl=400$ for the typical
bases size of $20 d \times 20 d$, to obtain approximately one layer of
particles) with the largest $z-$values. Bulk layers that were analyzed
included all particles whose centers were located within $0.5 d$ of
the plane located at $z=\zlayer$. This yielded $\Nl\approx 400-440$
particles in the layer, depending on $\theta$.  The results are not
sensitive to variations of $\Nl$ or bulk layer thickness by $10-15\%$.

\begin{figure}[!ht]
\includegraphics{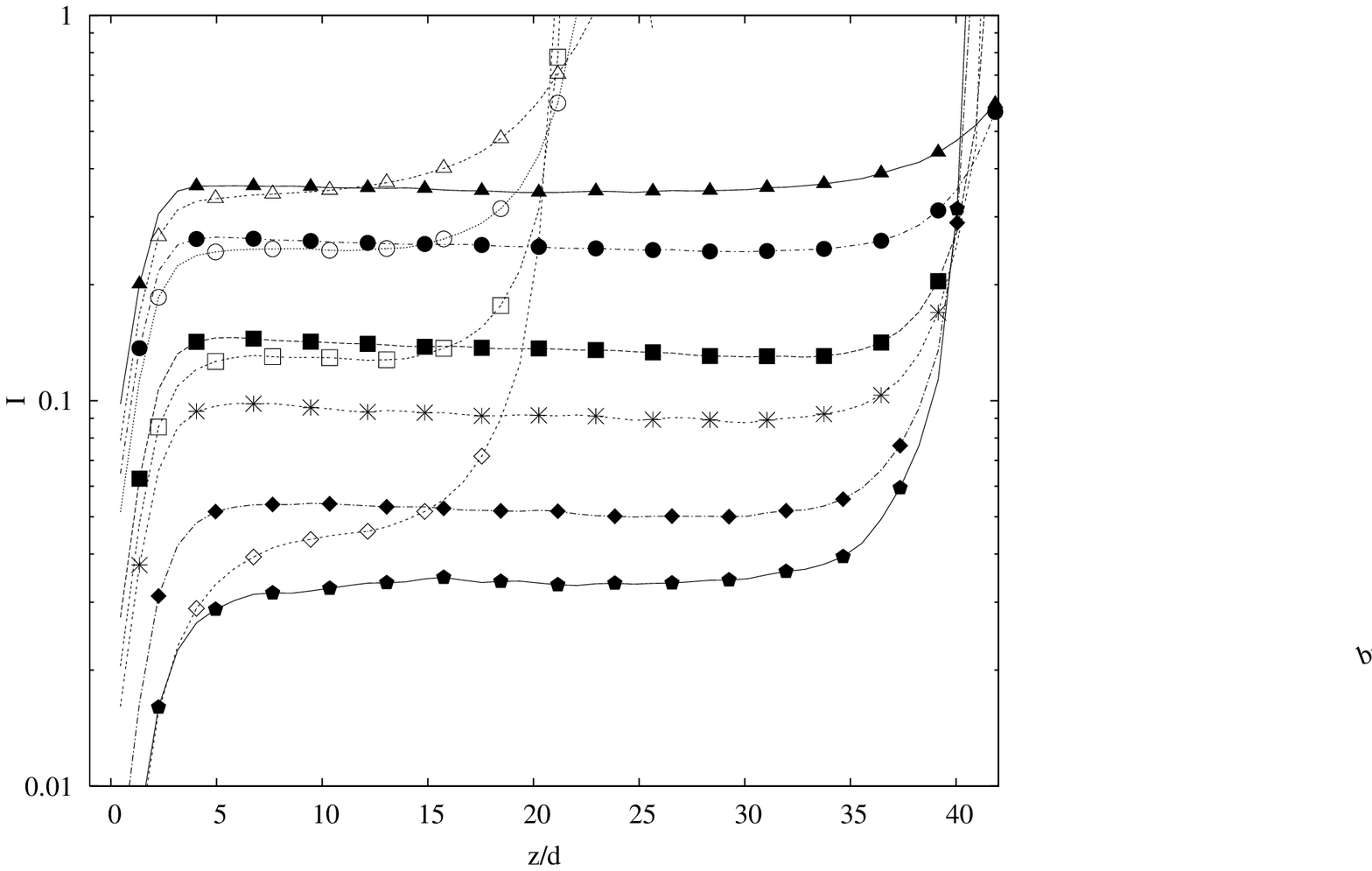}
\caption{\label{fig:prof_I} Scaling variable $I$ versus scaled height
$z/d$.}
\end{figure}
The values for $\zlayer$, reported in Table~\ref{tab:layers_part}, are
chosen near the center of the pile, in order to minimize boundary
effects from the top and bottom. In order to ensure that bulk
properties are observed in these layers, in Fig.~\ref{fig:prof_I} we
display the scaling variable $I$ [cf. Eq.(\ref{eq:gdrm})] as a
function of $z-$position, along with the position of the layers for
which detailed analysis is conducted. $I$ is expected to be constant
in the bulk~\cite{midi04}.

In the PIV technique used in laboratory experiments, velocities are
obtained by measuring particle displacements between two frames
separated by the time $\dt$. Similarly, we measure the velocity of
each particle $i \in \{1..\Nl\}$ in a layer from its displacement over
a characteristic ``measurement time'' $\dt$:
\begin{equation}
\vd^i (t)   = \left ( r_\alpha(t)-r_\alpha^i(t-\dt )\right ) /\dt.
\label{eq:vel}
\end{equation}
Here $\alpha=\{x,y,z\}$ is a coordinate label and ${\bf r}^i(t)$ is the 
position of particle $i$ at time $t$. 

Due to finite size effects, the instantaneous mean velocity of 
each layer,
\begin{equation}
\mv(t) = \sum_{i}^{\Nl}\vd^i(t) /\Nl,
\label{eq:mv}
\end{equation} 
fluctuates around the time-averaged velocity profile shown in
Fig.~\ref{fig:prof}(b-c). In order to avoid introducing spurious
correlation effects at large distances, we use the instantaneus mean
velocity to obtain the velocity fluctuations, rather than the
long-time mean velocity:
\begin{equation}
u_{\alpha}^i(t)   = \vd^i(t) - \mv(t).
\label{eq:ua}
\end{equation}
Finally, to obtain better statistics, we perform the correlation
analysis for $N_T\approx 10^4$ configurations separated by
$100\,\ts$. Although not explicitly indicated, all velocities $\vd^i$
have an implicit dependence on the measurement time $\dt$.

We compute six two-point velocity correlation functions in the
$xy-$plane for each layer, following the method in
Ref.~\cite{pouliquen04}.  Correlations in the $\beta=\{x,y,z\}$
component of the velocity at a distance $r$ along the $\alpha=\{x,y\}$
direction are given by
\begin{equation}
C_{\alpha\beta}(r)=\frac
{\sum_t [ \sum_{i,j} u_\beta^i u_\beta^j \prod_{\gamma=\{x,y\}}
\delta (r_\gamma^i - r_\gamma^j + \delta_{\gamma\alpha}r ) ]}
{\sum_t [ \sum_{i,j} \prod_{\gamma=x,y}
\delta (r_\gamma^i - r_\gamma^j + \delta_{\gamma\alpha}r )  ]}.
\label{eq:cxx}
\end{equation}
Here $\delta(r)=\frac{1}{\sqrt{2\pi w^2}} \exp(-\frac{r^2}{2w^2})$ is
a Gaussian function of width $w=0.4\, d$, as used in
Ref.~\cite{pouliquen04}, and $\delta_{\gamma\alpha}=1$ when $\gamma =
\alpha$ and $0$ otherwise. We do not examine correlations along the
$z-$direction since these connect particles with different average
velocities and strain rates, which could not be statistically averaged
in a satisfactorily manner.

In virtually all cases, a good fit to an exponential decay is
obtained. Nonexponential behavior was typically related to finite-size
effects, poor statistics or dispersion effects, which will be
discussed in Section~\ref{sec:surface}. A correlation length $\lab$ is
associated with each correlation function by obtaining the best linear
fit to a log-linear plot:
\begin{equation}
  C_{\alpha\beta}(r)= C_{\alpha\beta}(0)\exp (-r /\lab).
\label{eq:cab}
\end{equation}
Note that Pouliquen~\cite{pouliquen04} defined the correlation length
as the distance $L_{\alpha\beta}$ for which $\Cab (r) = 0.07 \Cab(0)$,
which is related to our correlation length through
\begin{equation}
\Lab\approx 2.66 \lab.
\end{equation}

\section{Results}

\subsection{\label{sec:surface}Velocity Correlations at the Surface}


\begin{figure}[!ht]
\includegraphics{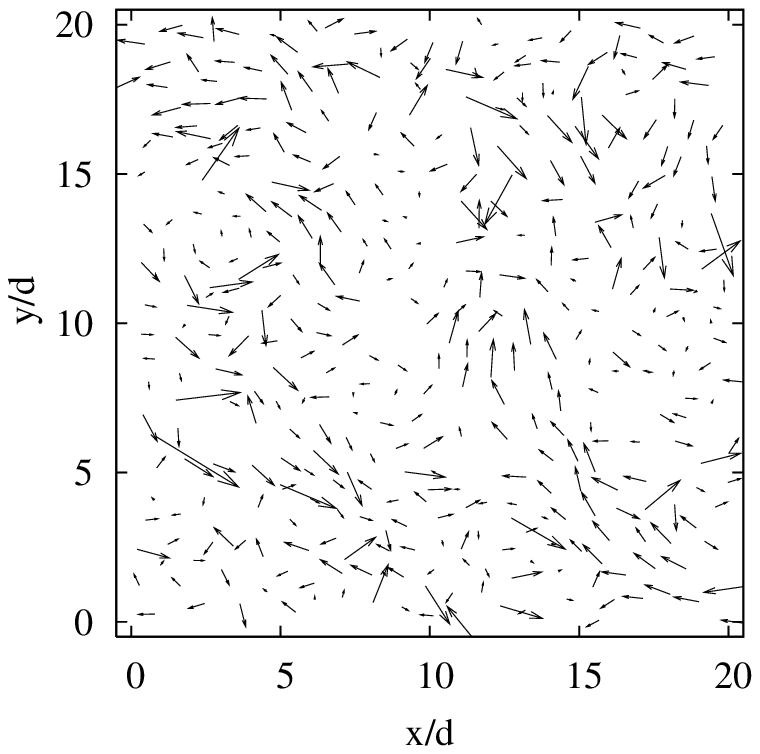}
\includegraphics{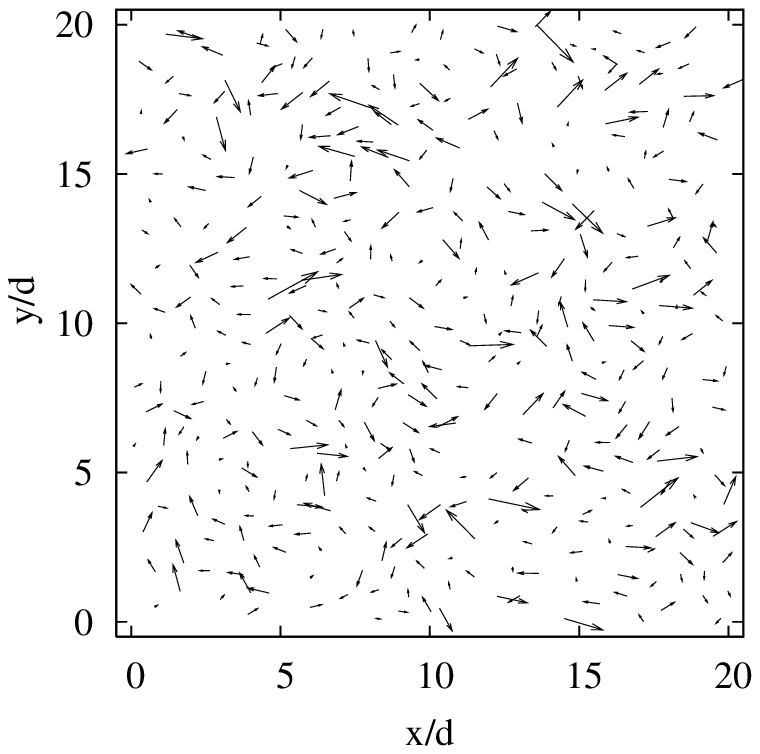}
\caption{\label{fig:vf} Sample velocity fields in the $xy-$plane for
the surface layers of two runs with $h/d \simeq 20$ and
$\dt/\tau_0 = 1$. (a) $\theta=21\dg$, (b) $\theta=23\dg$.}
\end{figure}

Figure \ref{fig:vf} shows sample vector plots of the velocity field
${\bf u}^i(t)$, projected on the $xy-$plane, with $\dt/\tau_0 = 1$,
for the surface layers of two simulation runs with $h/d \simeq 20$ and
incline angle $\theta=21\dg$ and $\theta=23\dg$ respectively. Visual
inspection of the patterns suggest more correlated flow structures for
the lower angle run, similar to observations made in
Ref.~\cite{pouliquen04}.

\begin{figure}[!ht]
\includegraphics{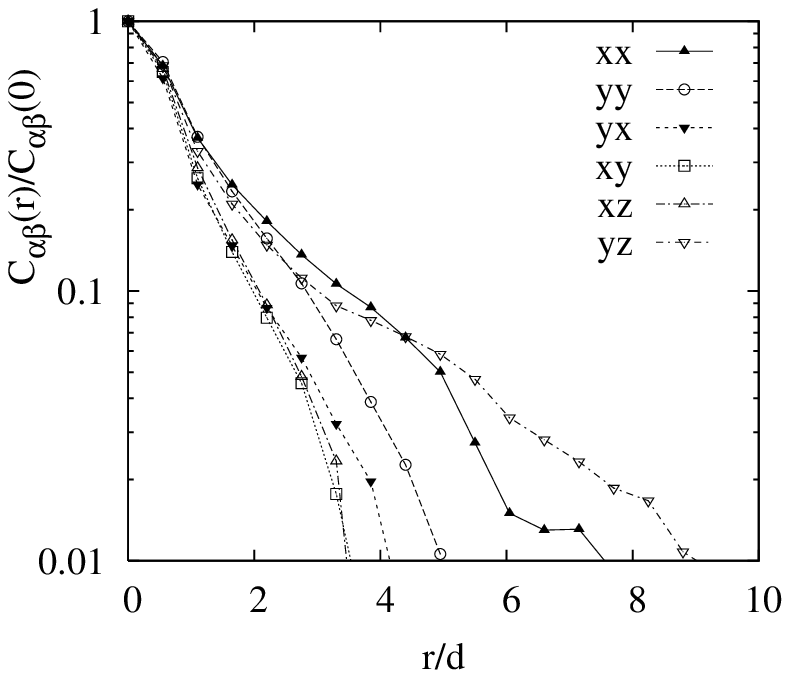}
\includegraphics{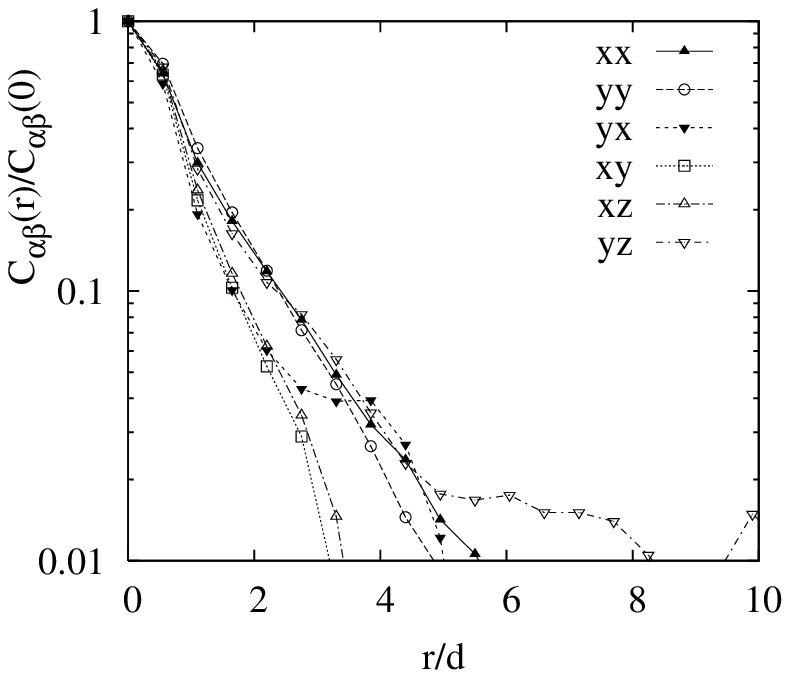}
\caption{\label{fig:surf} The six normalized correlation functions,
$\Cab (r)$, for two runs with $h/d \simeq 20$ and $\dt/\tau_0=1$.
(a) $\theta=21\dg$ and (b) $\theta=23\dg$.}
\end{figure}

\begin{figure}
\includegraphics{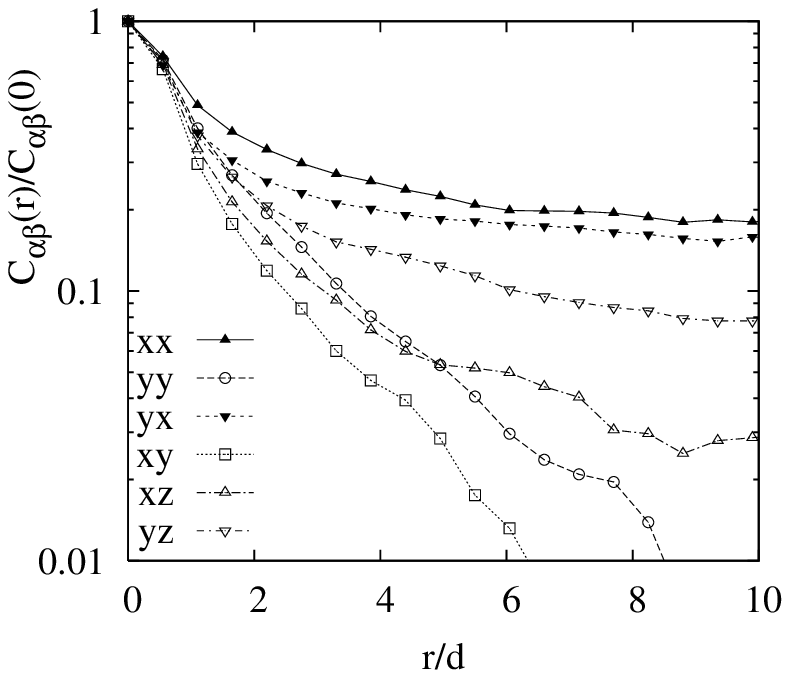}
\includegraphics{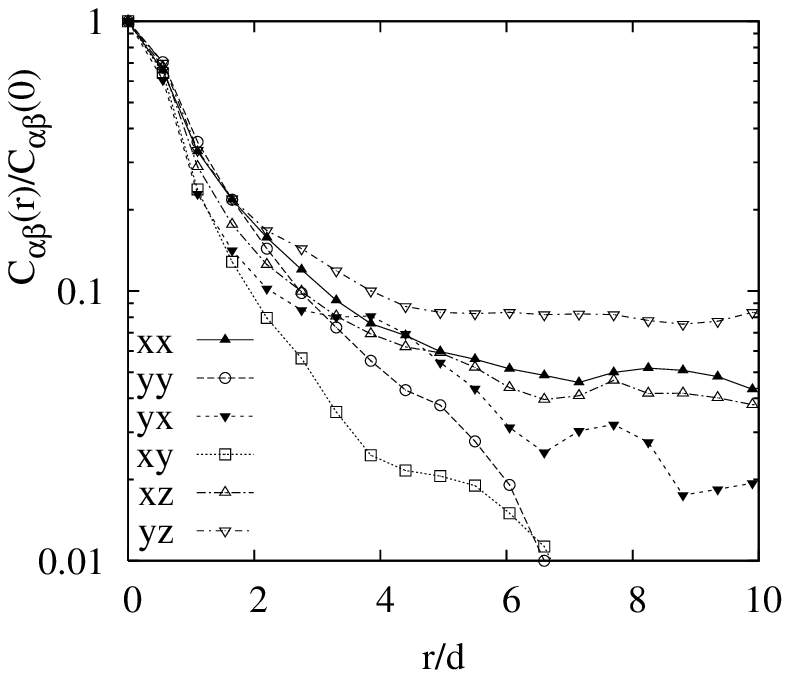}
\caption{\label{fig:surfgl} The six normalized correlation functions,
for the same runs shown in Fig~\ref{fig:surf}, but with relative
velocities computed by subtracting the time-averaged mean velocity 
instead of the instantaneous mean velocity. Note the ``flattening''
of the correlation functions at large distances (see text).}
\end{figure}

For a more quantitative analysis of the velocity correlations, in
Fig.~\ref{fig:surf} we show all six normalized correlation functions
(cf. preceding Section) for the pair of runs discussed above.  From
the linear decrease of $\log[\Cab (r)]$ with distance, we can
determine a correlation length with each normalized correlation
function by fitting to Eq.(\ref{eq:cab}).

\begin{figure}[!ht]
\includegraphics{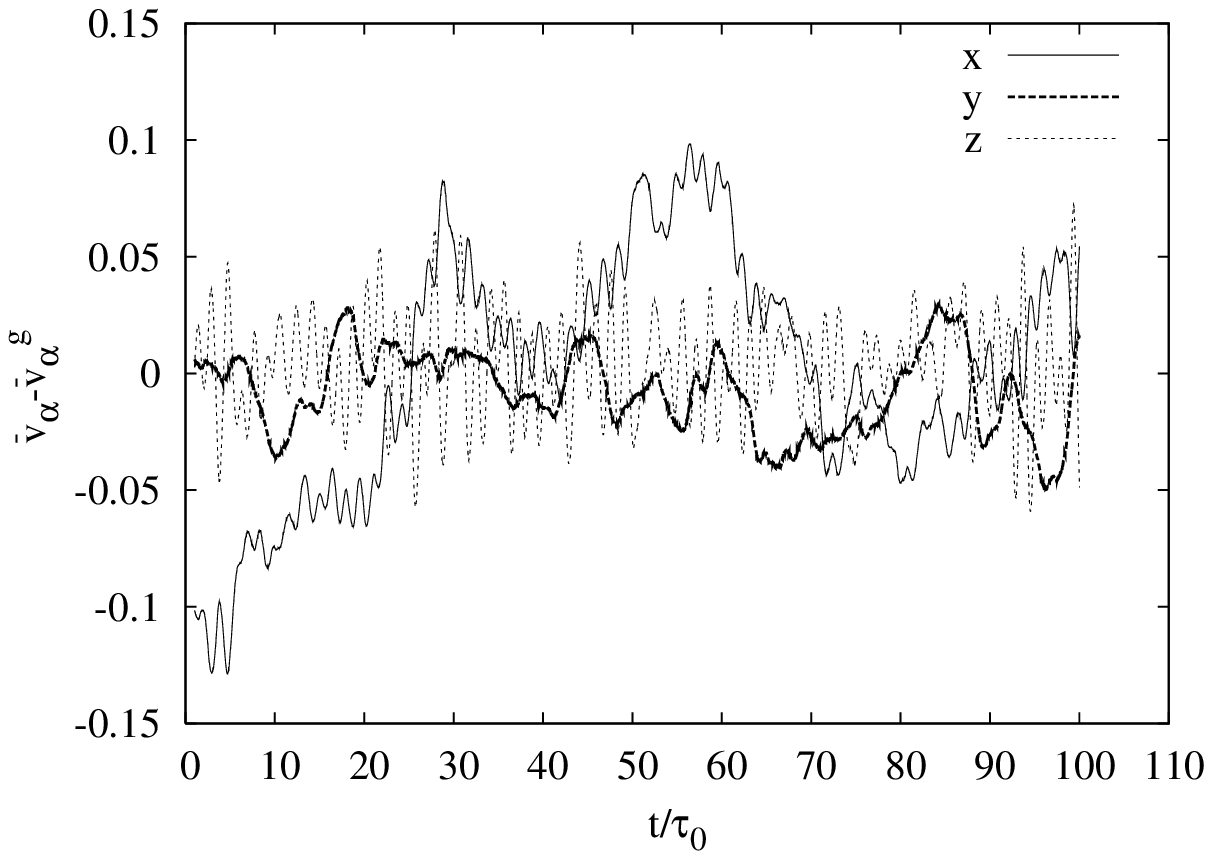}
\caption{\label{fig:frame_surf} Fluctuating components of surface
layer velocity for $h/d \simeq 20$, $\theta=21\dg$.}
\end{figure}
It is instructive to compare the profiles in Fig.~\ref{fig:surf} to
those in Fig.~\ref{fig:surfgl}, which are obtained from exactly the
same data set, but by using the time-averaged mean velocity instead of
the instantaneous mean velocity to compute velocity fluctuations.
Effectively, this corresponds to switching the order of spatial
averaging and temporal averaging and is not expected to affect the
result in a large system in steady-state.  However, spurious
correlations are obtained when the latter method is used, which are
due to the fluctuations in the instantaneous mean velocity shown in
Fig.~\ref{fig:frame_surf}. The largest deviations in
Fig.~\ref{fig:surfgl}(a) arise for $C_{\alpha x}(r)$, consistent with
the largest deviations in $\bar{v}_{x}(t)$ seen in
Fig.~\ref{fig:frame_surf}.  The resemblance of these to some of the
correlation functions shown in Ref.~\cite{pouliquen04}, which uses the
global mean velocity, suggests that the combination of the observed
non-exponential tails and the particular definition of the correlation
length $\Lab$ used may have resulted in over-estimation of the
reported correlation lengths. We typically obtain the least systematic
bias in the $y-$component of the velocity and therefore rely mainly on
$\cyy (r)$ for comparative analysis of runs.

\begin{figure}[!ht]
\includegraphics{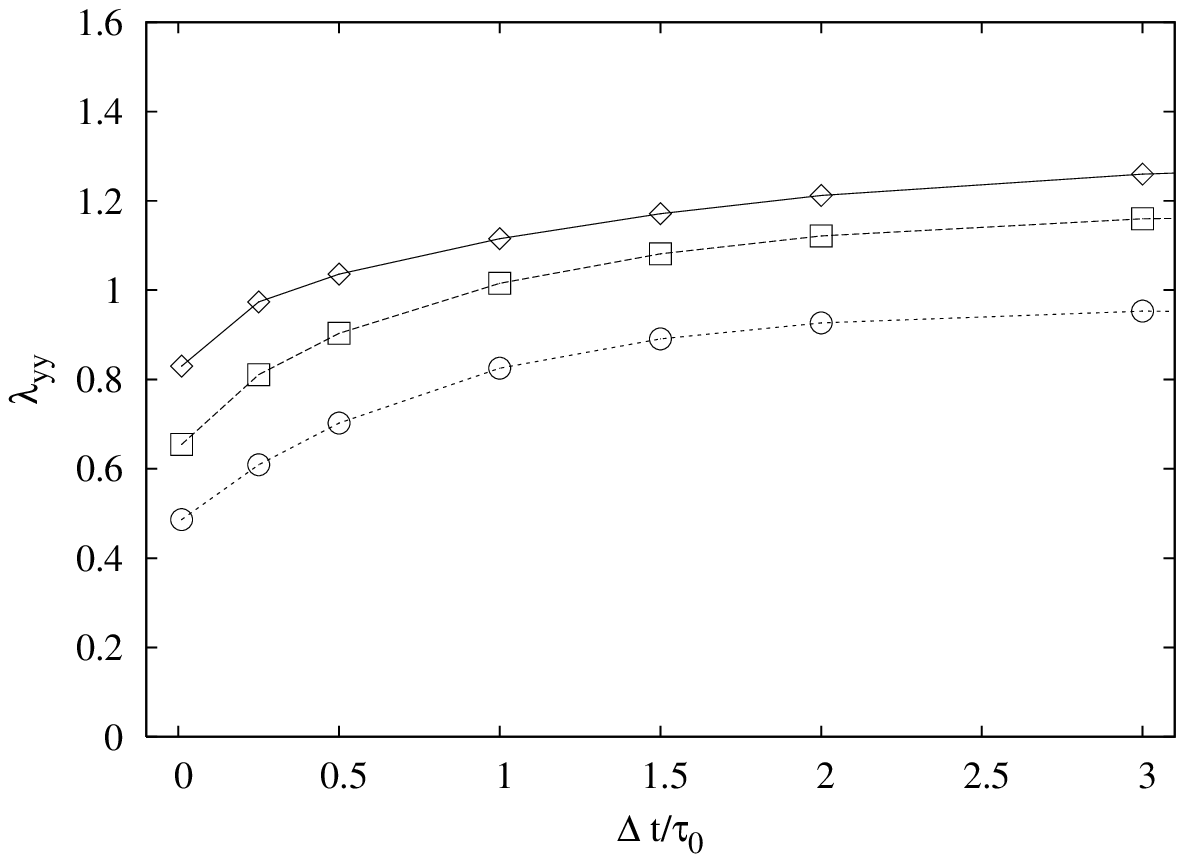}
\caption{\label{fig:lyy_surf} Surface layer correlation length as a
function of $\dt$ for $h/d \simeq 20$ and $\theta = 21\dg$
($\lozenge$), $\theta = 23\dg$ ($\square$) and $\theta = 25\dg$
($\bigcirc$).}
\end{figure}
Although we confirm that the measured correlation is indeed larger for
the smaller angle run, we find that the velocity correlation lengths
depend on the measurement time, $\dt$.  As seen in
Fig.~\ref{fig:lyy_surf}, the correlation length initially increases
with $\dt$, ultimately reaching a maximum value as the mean strain
experienced by the particles during the measurement time,
$\epsilon\equiv \dot\gamma \dt$, exceeds 1.  The plateau at large
$\dt$ is understandable: The velocity correlations are naturally
disappearing as particles start to execute diffusive motion induced by
the strain. However, the initial increase in $\lyy$ makes it difficult
to decide the ``correct'' measurement time to use in order to
appropriately compare runs with different heights and angles. We
discuss this issue in more detail in Appendix~\ref{sec:fluc} and
arrive at the conclusion that the results should be compared at the
fixed values of measurement strain $\epsilon=\dot\gamma \dt$.

\subsection{\label{sec:depth}Depth dependence of velocity correlations}

Since our main interest is in the nature of granular flow in the bulk,
we next investigate how the velocity correlation length varies as a
function of distance from the surface by taking advantage of the
information available in simulations but usually not in experiments.

\begin{figure}[!ht]
\includegraphics{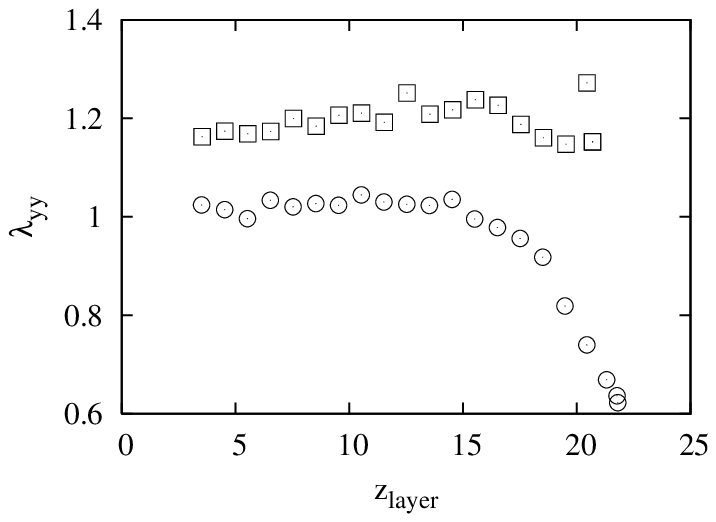}
\includegraphics{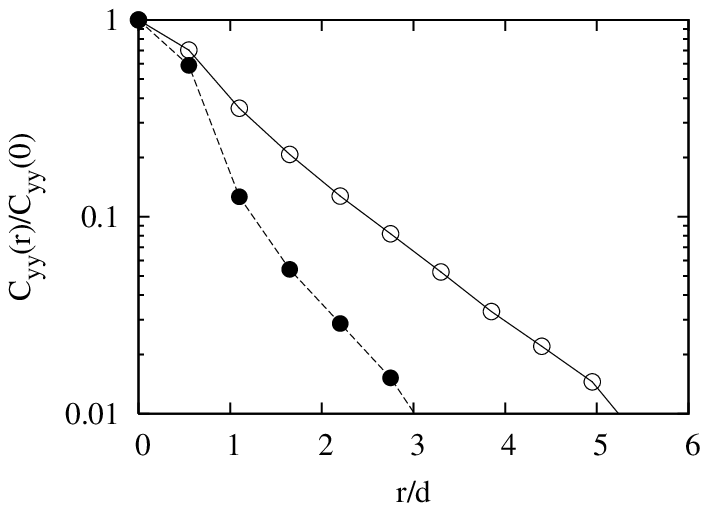}
\caption{\label{fig:depth} (a) Correlation length $\lambda_{yy}$
  profiles for angles $23\dg$ (squares) and $25\dg$ (circles), height
  $h/d \simeq 20$. (b) $C_{yy}(r)$ at the surface (filled circles) and in the
  bulk (open circles) for $\theta = 25\dg$ and $h/d \simeq 20$. All results
  are obtained for fixed strain $\epsilon = 0.1$.}
\end{figure}
Figure \ref{fig:depth}(a) shows the correlation length profiles as a
function of interrogated layer position $\zlayer$ for $h/d \simeq 20$
and angles $23\dg$ and $25\dg$. The correlation length is slightly
smaller near the top of the pile compared to the bulk, where it is
approximately constant. The correlation lengths increase as the angle
is decreased towards the angle of repose.  This is consistent with
earlier observations reported on surface
correlations~\cite{pouliquen04}.

Figure~\ref{fig:depth}(b) shows typical shapes of correlation
functions for $h/d \simeq 20$, $\theta=25\dg$, $\epsilon = 0.1$ in the bulk
compared to the surface. $C_{yy}(r)$ on the surface has more noise
than in the bulk, which is most probably due to the saltating
particles. Fluctuations in the mean velocity are dominated by a
relatively small number of saltating particles at the surface.  Since
such particles do not exist in the bulk layers, the statistics is
better in the bulk layers. Furthermore, the rheology of chute flow
becomes more robust and better characterized in the bulk. Therefore,
in order to make a more quantitative connection between the
correlation length $\lyy$ and deposit function $\hstop(\theta)$, we
will focus on the bulk flow for the remainder of this paper.

\subsection{\label{sec:bulk} Correlations in the bulk}

\begin{figure}[!ht]
\includegraphics{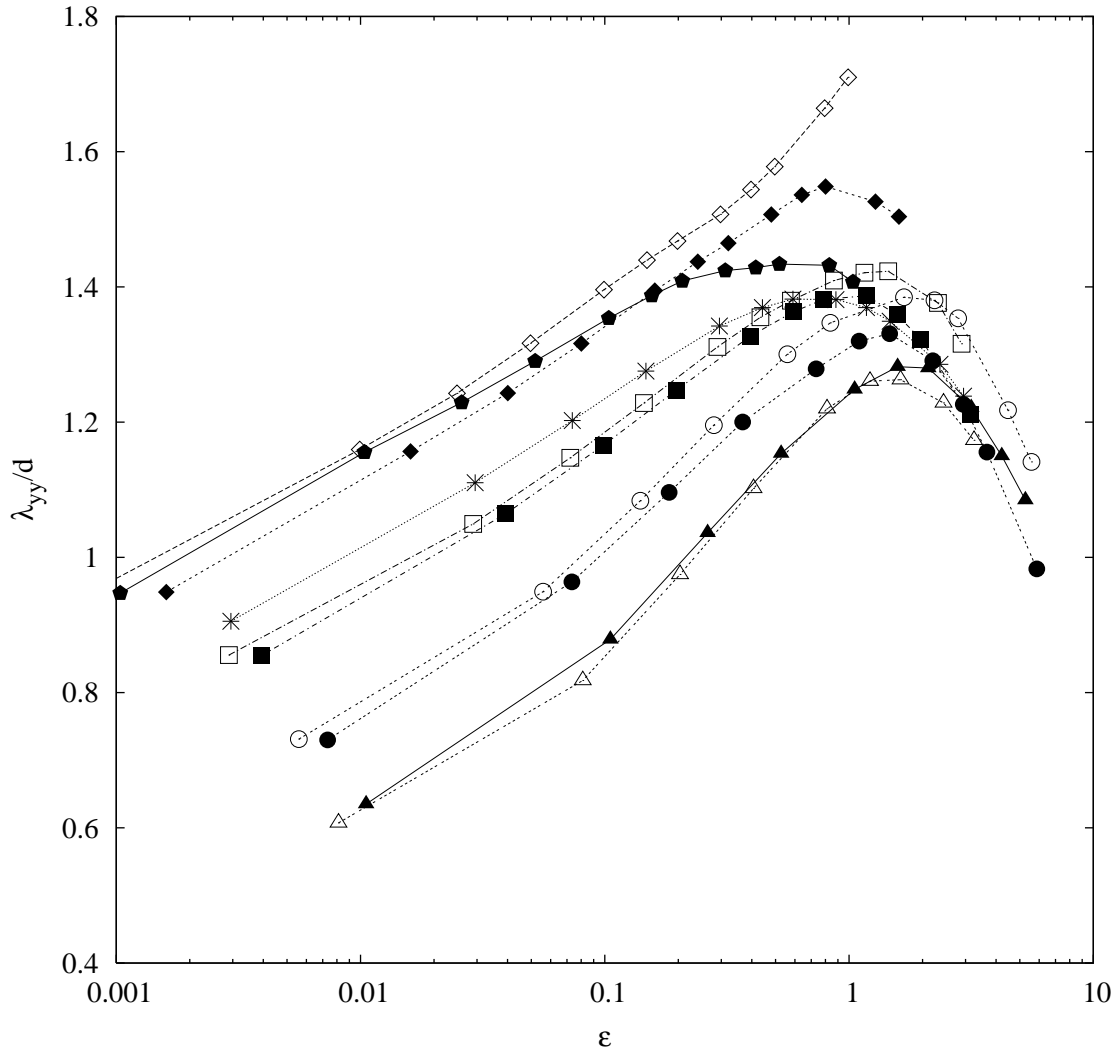}
\caption{\label{fig:lyy_epsilon} Correlation length $\lambda_{yy}$ as
a function of strain.}
\end{figure}

Figure~\ref{fig:lyy_epsilon} shows the correlation length
$\lambda_{yy}$ as a function of $\epsilon$, for all the runs described
in Table~\ref{tab:layers_part}. As discussed is
Section~\ref{sec:surface}, the comparison of correlation lengths at
different angles should be done at a fixed value of $\epsilon$, for
which the rattling motion has largely averaged away and decorrelation
due to the diffusive motion has not yet set in. For the systems listed
in Table~\ref{tab:layers_part}, the region $0.1 < \epsilon < 0.2$ is
the most suitable.

\begin{figure}[!ht]
\includegraphics{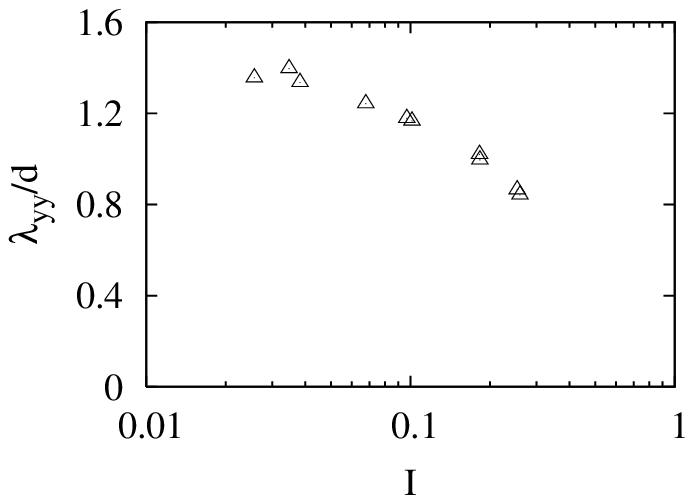}
\includegraphics{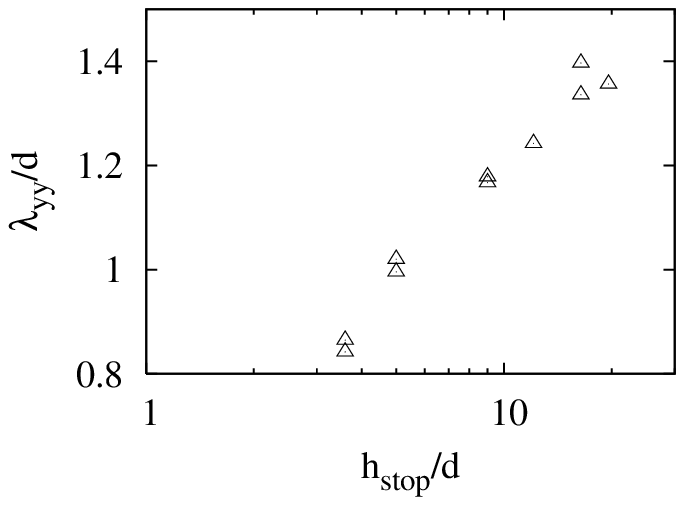}
\caption{\label{fig:lyy_e0.3} (a) $\lambda_{yy}$ versus $I$ and (b)
$\lambda_{yy}$ versus $\hstop$ for $\epsilon=0.1$.}
\end{figure}

In Fig.~\ref{fig:lyy_e0.3}(a), $\lyy$, measured at $\epsilon = 0.1$ as
a function of $I$, is displayed. The dependence is quite weak, and
$\lyy$ appears to diverge only logarithmically, if at all, in the
$I\to 0 $ limit. In contrast, $\hstop$ exhibits stronger dependence
and power-law divergence in the same limit. The interdependence of
$\lyy$ and $\hstop$ is directly probed in Fig.~\ref{fig:lyy_e0.3}(b),
which reveals an approximate relationship $\lyy/d \sim
\ln(\hstop/d)$. This result suggests that the velocity correlation
length remains much smaller than the pile height for all flowing piles
in the $I\to 0 $ limit and does not directly influence $\hstop$.  In
fact, the situation appears to bear an interesting similarity to that
in molecular glasses, where the diverging viscosity at the glass
transition is accompanied by a region of cooperative motion that
remains quite small and appears to show signs of a logarithmic
divergence~\cite{bouchaud}.

To probe smaller values of $I$ one needs to study thick piles at low
angles just above the angle of repose. However, it has recently been
shown by Silbert~\cite{silbert05} that these flows exhibit flow
instabilities. For example, for ($h\simeq 80$) and $\theta \le 21\dg$,
a resonant normal vibration mode is present which results in a
longitudinal dilation wave in the $z-$ direction.  As a result, the
density in a given layer, as well as the height of the flow, varies
periodically in time, breaking the time-translation symmetry.

\section{\label{sec:conclusion} Conclusions}
We have presented numerical evidence that the spatial correlations
associated with velocity fluctuations in chute flow decay
exponentially with distance and remain short-ranged, with the
correlation lengths exhibiting at best a logarithmic divergence as the
angle of repose is approached. In order to make meaningful comparisons
of correlation lengths, it is important to choose measurement times
that are inversely proportional to the local strain rate. As described
in Appendix~\ref{sec:fluc}, the analysis of single particle motion
reveals two superimposed motions: Cage rattling that dominates the
instantaneous velocity fluctuations but lacks correlation and
sterically constrained motion that causes strain-induced diffusion and
reflects correlations in particle rearragement.  A closed-form set of
rheological equations would need to relate hydrodynamic flow
parameters to the microscopic energy dissipation in the system, and
unraveling this relationship would likely need to take into account
the quite different dissipation characteristics of these two types of
motions.

\begin{acknowledgments}
Sandia is a multiprogram laboratory operated by Sandia Corporation, a
Lockheed Martin Company, for the United States Department of Energy's
National Nuclear Security Administration under Contract
No. DE-AC04-94AL85000.
\end{acknowledgments}

\appendix
\section{\label{sec:fluc} Dependence on effective strain}

\begin{figure}[!ht]
\includegraphics{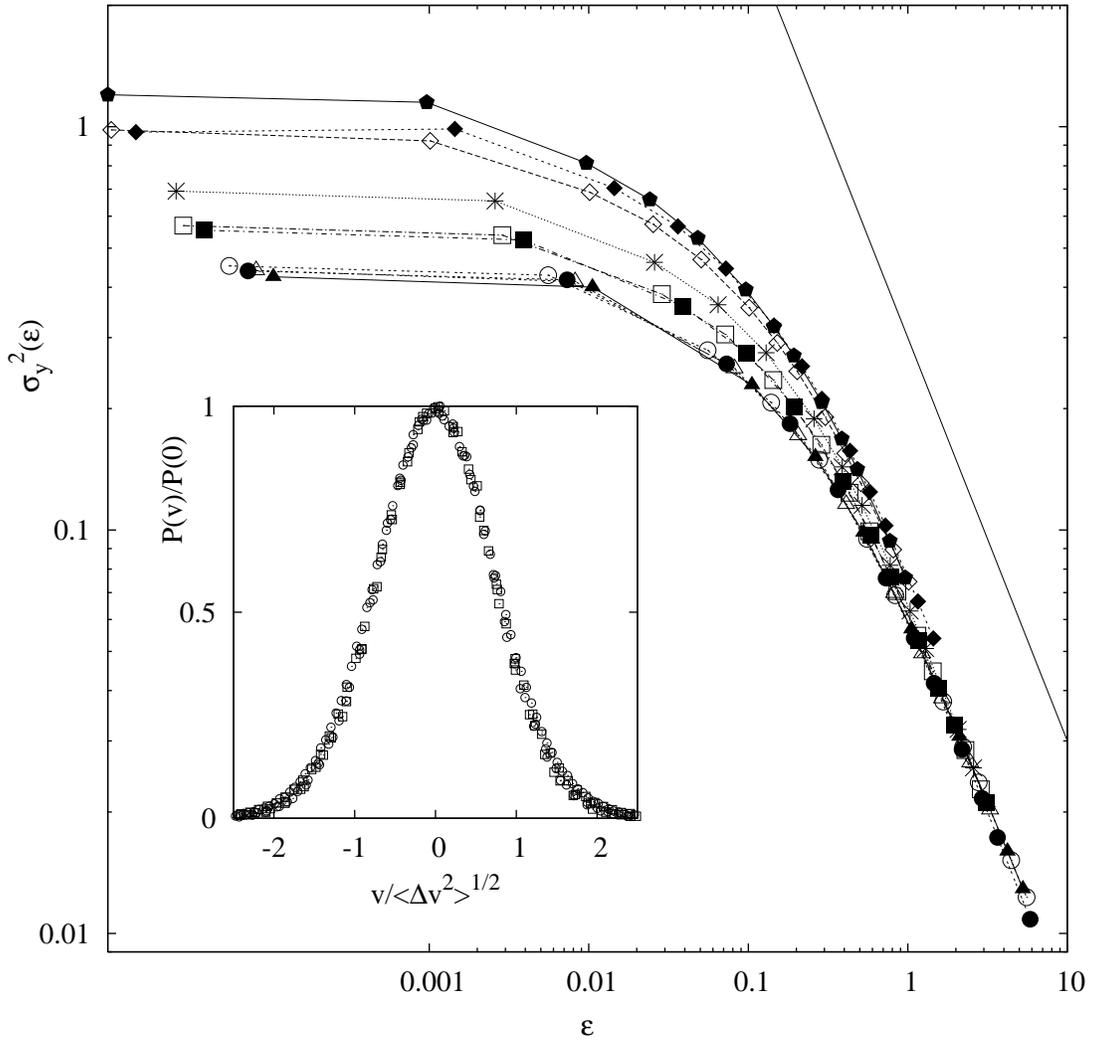}
\caption{\label{fig:cyy0_epsilon} $y$-component of the mean square
  velocity fluctuations versus strain, measured in bulk layers as
  characterized in Table~\ref{tab:layers_part}. Legend is shown
  in Table~\ref{tab:layers_part}. Inset shows the probability
  distribution functions for the y-velocity component in the bulk
  layer for different runs and $\dt$'s for $h/d \simeq 20$. The probability
  is scaled by maximum value and velocities scaled by variance.  }
\end{figure}

The velocity correlation results depend on measurement time, $\Delta
t$, and measurement strain $\epsilon=\dot\gamma \dt$. We found it
instructive to study the distribution of velocity fluctuations as a
function of measurement strain $\epsilon=\dot\gamma \dt$. To avoid the
effects of boundaries we show here the results for bulk layers listed
in Table~\ref{tab:layers_part}.  Figure \ref{fig:cyy0_epsilon} depicts
non-dimensionalized mean square velocity fluctuations as a function of
$\epsilon$:
\begin{equation}
\sigma_{\alpha}^2(\epsilon)\equiv\frac{\av{u_{\alpha}^2}}{(\dot\gamma d)^2},
\end{equation}
where $u_\alpha$ is obtained using a measurement time 
$\dt = \epsilon/\dot\gamma$.

The inset in the figure shows that a simple Gaussian distribution is
observed in all cases, such that the variance completely characterizes
the fluctuations. For an individual data set, we observe a
ballistic-like regime for small $\epsilon$, in which the measured
velocities do not depend on $\epsilon$, whereas particles exhibit
diffusive motion at larger values of $\epsilon$.

The collapse of all data for large $\epsilon$ suggests that the
diffusive motion at long times is dictated by steric constraints
between particles as they pass near each other, such that the
displacement of each particle depends only on accumulated strain and
not the strain rate. Hence, the velocity fluctuations arising from
this motion do not depend on the scaling variable $I$.

\begin{figure}[!ht]
\includegraphics{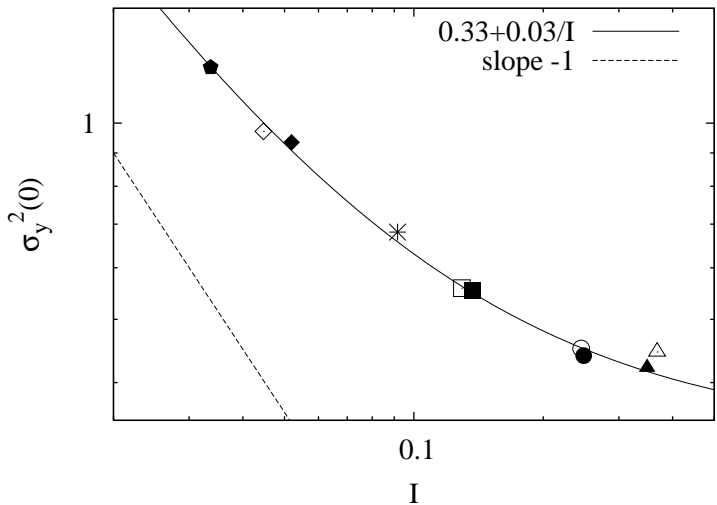}
\caption{\label{fig:vyy_I} Velocity fluctuations versus inertial
  number $I$ for the systems listed in Table~\ref{tab:layers_part}.}
\end{figure}

In the ballistic-like regime the data-sets diverge from each other. In
particular, the instantaneous velocity fluctuations for small
$\epsilon$ exhibit anomalous scaling $\sigma_{\alpha}^2(0)\sim I^{-1}$
in the limit $I \to 0$, as shown in Fig.~\ref{fig:vyy_I} and observed
previously~\cite{midi04}.  We associate this additional motion at
short times with the rattling of particles in their temporary cages.  A
more quantitative description of this superimposed grain motion will
be discussed elsewhere~\cite{coherent}.  This motion is expected to
have very weak spatial correlation and average out quickly with
increasing measurement time, while still dominating the instantaneous
velocity fluctuations and giving rise to their anomalous scaling.
Since the overall motion of each particle is the superposition of
these two motions, it is not surprising that the measured correlation
length is suppressed at small measurement times.  Since we are
interested in identifying the nature of correlations associated with
the sterically hindered motion of the particles, we compare
correlation lengths for different systems at a fixed value of
measurement strain that is large enough to have substantially
eliminated the effects of cage motion, but small enough that the
diffusive motion has not caused substantial decorrelation. Note that
this effectively changes the measurement time that must be used to
determine velocities according to the local strain rate.

\section{\label{sec:large} Effect of system size and particle's stiffness}

\begin{figure}[!ht]
\includegraphics{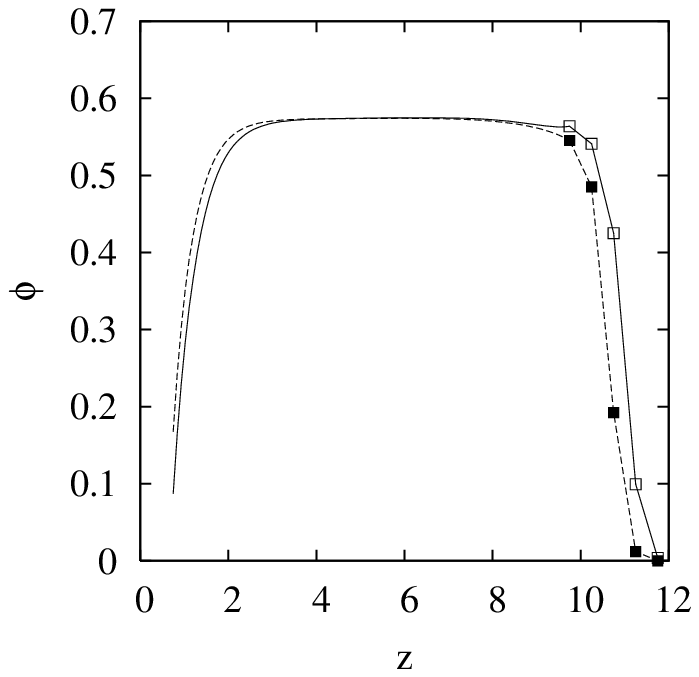}
\includegraphics{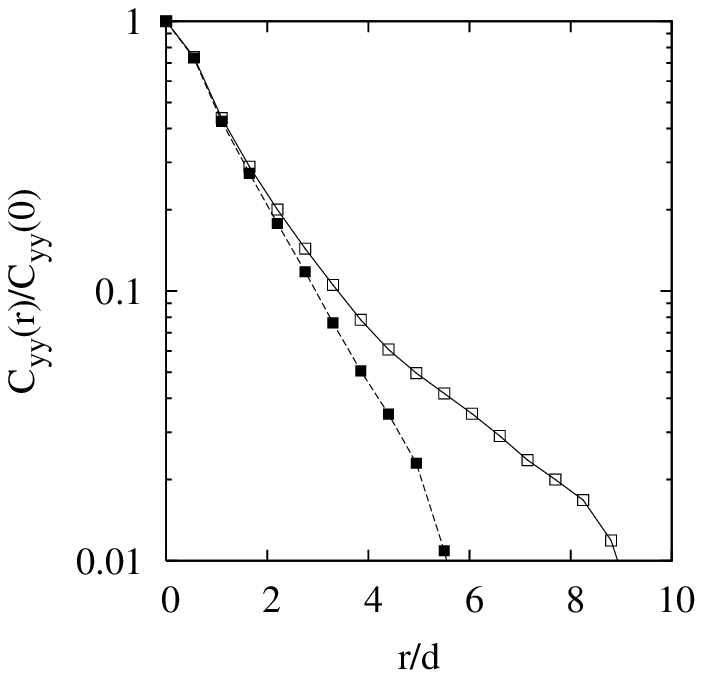}
\caption{\label{fig:vc_n18} Volume fraction profiles (time-averaged
  and smoothed) (a) and velocity
correlation functions (b) for $20d \times 20d$ ($\square$) and $40d
\times 40d$ ($\blacksquare$) base area with $h/d \simeq 10$, $\theta=23\dg$,
$\dt /\tau_0 = 1 $.}
\end{figure}
To test the system size effect on the correlation length we
equilibrated additional configurations of height $h/d \simeq 10$ at
$\theta = 23\dg$ with two sizes of base area, $20d \times
20d$ and $40d \times 40d$.

Figure~\ref{fig:vc_n18}(a) shows that the volume faction profiles for
the system with base area $40d \times 40d$ is the same as for typical
configurations whose bases are $20d \times 20d$.
Figure~\ref{fig:vc_n18}(b) shows the velocity correlation functions
for the large system compared with the small system size. The
correlation lengths, obtained from the exponential fit to the first
four points, are very similar, $L_{yy}(40\times 40)=1.295$,
$L_{yy}(20\times 20)=1.240$ , $L_{xx}(40\times 40)=1.194$,
$L_{xx}(20\times 20)=1.141$, (less then 5\% difference). The deviation
in exponential behavior in the smaller system is larger. The first
four points in correlation function are, however, the same. This
justifies our method of obtaining $\lambda_{yy}$ by fitting the
exponential decay function to the first four points of correlation
data.

\begin{figure}[!ht]
\includegraphics{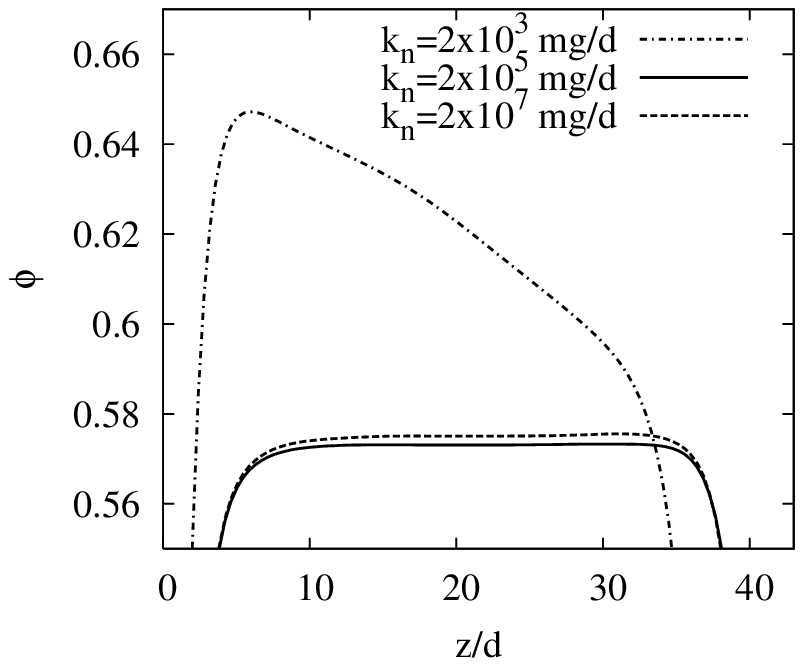}
\includegraphics{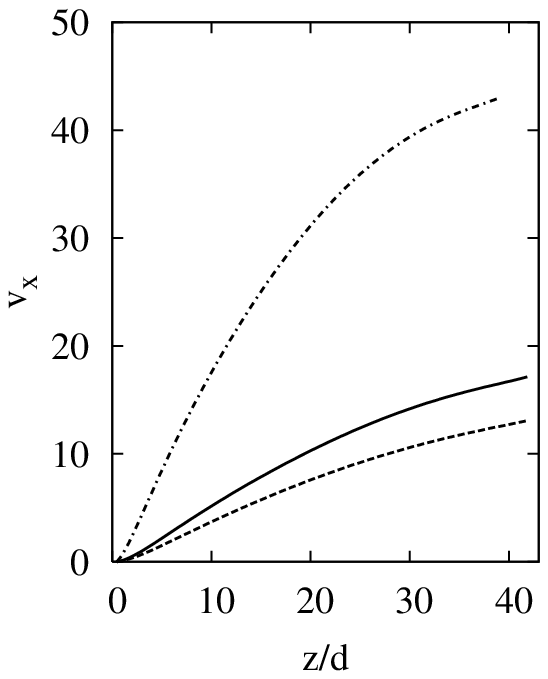}
\caption{\label{fig:vc_n17.9s} (a) Volume fraction profiles for
  different values of grain stiffness (data smoothing routine used in
  addition to time averaging). $h/d \simeq 40$ and $\theta =
  23\dg$.(b) Velocity profiles for the same configurations shown in
  (a).}
\end{figure}

\begin{figure}[!ht]
\includegraphics{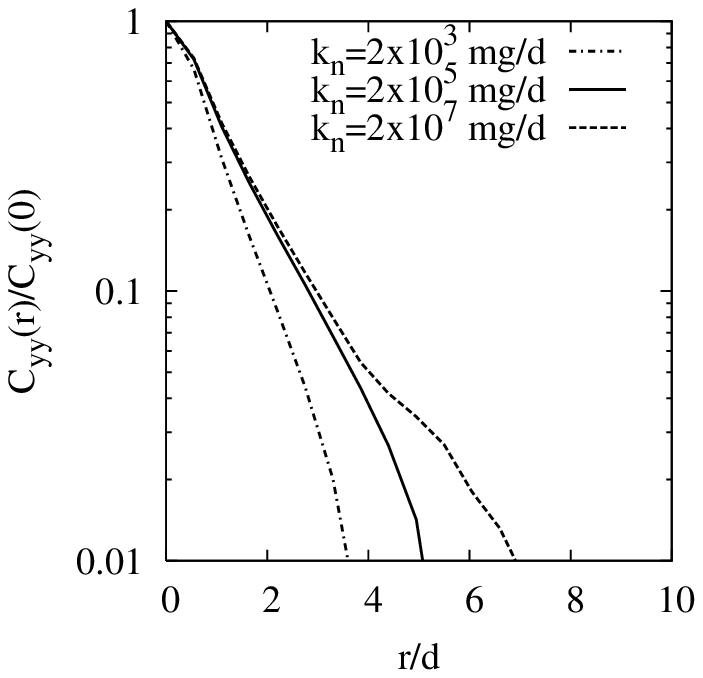}
\includegraphics{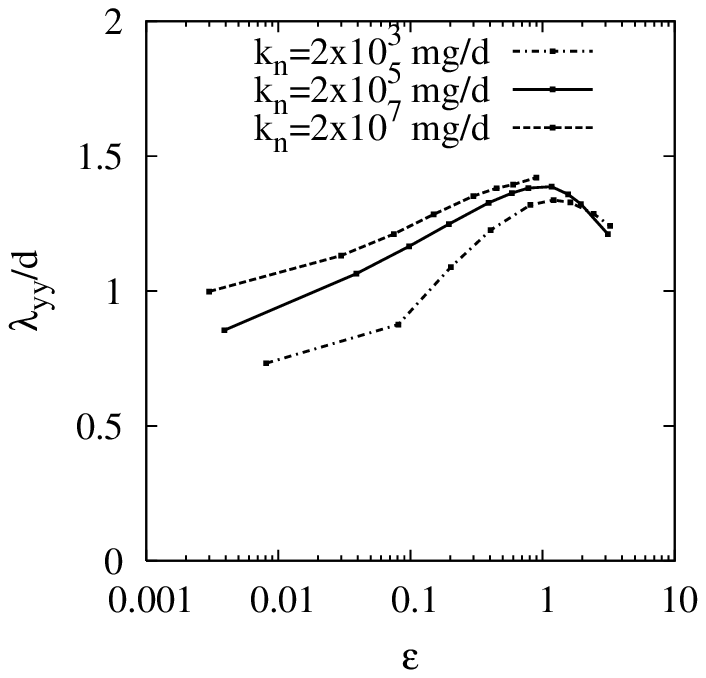}
\caption{\label{fig:vc_n17.9s2} (a) Velocity correlation function at
  $\epsilon \simeq 0.1$ and (b) correlation lengths for three
  values of grain stiffnes. $h/d \simeq 40$ and $\theta = 23\dg$. }
\end{figure}

To study the dependence on spring stiffness $k_n$, we carried out
simulations for the tall pile $h/d \simeq 40$ at $\theta=23\dg$ with
$k_n=2\times 10^3 \, mg/d$ and $k_n=2\times 10^7 \, mg/d$ and compared
the results with earlier $k_n=2\times 10^5 \, mg/d$ results. Changes
of stiffness affect the collision time and restitution coefficients
and require appropriate adjustments of the timestep and viscoelastic
parameter~\cite{silbert01}. Therefore, for the case of increased
stiffness, $k_n=2\times 10^7 \, mg/d$, the timestep was decreased to
$\ts/\tau_0=10^{-5}$ and $\gamma_n$ was set to $\gamma_n=500 \,
\sqrt{g/d}$. In the case of the soft material, $k_n=2\times 10^3 \,
mg/d$, the timestep was kept at the same typical value,
$\ts/\tau_0=10^{-4}$, and the viscoelastic dumping constant was
reduced to $\gamma_n=5 \, \sqrt{g/d}$.

Figure~\ref{fig:vc_n17.9s}(a) shows that increasing the stiffness by
two orders of magnitude increases the value of bulk volume fraction
from $0.573$ to only $0.575$. The stiffer grains flow with slightly
lower velocities, Fig.~\ref{fig:vc_n17.9s}(b). Decreasing the
stiffness by two orders of magnitude has a big qualitative effect. The
bulk is no longer characterized by a constant volume fraction. The
grains closer to the bottom are compressed by the weight of the
system, leading to a decreasing with depth volume fraction profile.
Figure~\ref{fig:vc_n17.9s2}(a) shows the effect of the coefficient of
stiffness on the two-point correlations function $C_{yy}(r)$, measured
at $\dt/\tau_0=0.1$ for soft grains and $\dt/\tau_0=0.25$ for typical
and stiff grains. This choice of $\dt$ provides approximately equal
value of measurement strain, $\epsilon \simeq 0.1$.  Again, we note a
negligibly small difference between the results with typical and stiff
grains. The results for soft grains indicate a smaller correlation
length. The difference between correlation length for the three
configurations is small and decreases with the increase of $\epsilon$,
see Fig.~\ref{fig:vc_n17.9s2}(b). In summary, the effect of variations
in $k_n$ is minimal as long as $k_n \ge 2\times 10^5 \, mg/d$.


\end{document}